\documentclass[epj]{svjour}

\usepackage{graphicx}

\begin{document}

\newcommand{\ep}{\varepsilon }
\def\sl#1{\slash{\hspace{-0.2 truecm}#1}}
\def\beqn{\begin{eqnarray}}
\def\eeqn{\end{eqnarray}}
\def\nn{\nonumber}

\title{Single spin asymmetries in elastic electron-nucleon scattering 
}
\author{B. Pasquini\inst{1,2} \and M. Vanderhaeghen\inst{3,4}
}

\institute{ Dipartimento di Fisica Nucleare e Teorica, Universit\`a degli
Studi di Pavia and INFN, Sezione di Pavia, Pavia, Italy\and
ECT$^*$, Villazzano (Trento), Italy \and 
Thomas Jefferson National Accelerator Facility,
Newport News, VA 23606, USA, \and
Department of Physics, College of William and Mary,
Williamsburg, VA 23187, USA}
\date{Received: date / Revised version: date}
%
\abstract{
We discuss the target and beam normal spin asymmetries in 
elastic electron-nucleon scattering which depend on the imaginary part 
of two-photon exchange processes between electron and nucleon.
In particular, we estimate these transverse spin asymmetries 
for beam energies below 2 GeV, where the
two-photon exchange process is dominated from the resonance contribution
to the doubly 
virtual Compton scattering tensor on the nucleon. 
} 
\maketitle
\section{Introduction}
\label{intro}

Elastic electron-nucleon scattering in the one-photon exchange approximation 
is a valuable tool to access information on the structure of hadrons.
New experimental techniques exploiting polarization observables have made
possible precision measurements of hadron structure quantities, such as
its electroweak form factors, parity violating effects,
$N \to \Delta$ transition form factors, and
spin dependent structure functions. 
However, to push the precision frontier further in electron scattering, 
one needs a good control of $2 \gamma$ exchange mechanisms 
and needs to understand how they may affect 
different observables. 
The imaginary (absorptive) part of the $2 \gamma$ exchange amplitude 
can be accessed through a single spin asymmetry (SSA) in 
elastic electron-nucleon scattering, when either the target or beam spin 
are polarized normal to the scattering plane.
As time reversal invariance forces 
this SSA to vanish for one-photon exchange, it 
is of order $\alpha = e^2 / (4 \pi) \simeq 1/ 137$. Furthermore, to polarize   
an ultra-relativistic particle in the direction 
normal to its momentum involves 
a suppression factor $m / E$ (with $m$ the mass and $E$ the 
energy of the particle), 
which typically is of order $10^{-4} - 10^{-3}$ when the electron 
beam energy is in the 1 GeV range. Therefore, the 
resulting target normal SSA can be expected to be of order $10^{-2}$, 
whereas the beam normal SSA is of order $10^{-6} - 10^{-5}$.
In the case of a polarized lepton beam, asymmetries of the order ppm 
are currently accessible in parity violation (PV) elastic electron-nucleon 
scattering experiments. 
While the PV asymmetry measurements involve a beam spin polarized
along its momentum, the SSA for an electron
beam spin normal to the scattering plane can be accessed using the
same experimental apparatus.
Results from first measurements of this beam normal SSA 
have been presented in this conference.
\newline
\indent
Model calculations for such observables have been recently performed
in different kinematical regimes~\cite{AAM02,AM04,GGV04,DM04}.
Here we report a study of the imaginary part of the $2 \gamma$ exchange 
entering in the normal SSA's at low and intermediate beam energies~\cite{PV04}.
Using unitarity, one can relate the imaginary part of the $2 \gamma$  
amplitude to the electroabsorption amplitudes on a nucleon. 
Below or around two-pion production threshold, 
one is in a regime where these electroproduction amplitudes are relatively 
well known using pion electroproduction experiments as input. 
Therefore the aim is to gain a good knowledge of the imaginary part 
of the 
two-photon exchange amplitude, and then to use such information as input
for dispersion relations which will allow to quantify the contribution 
of the real part of the $2\gamma$ exchange processes.
In addition, 
observables such as normal SSA's are sensitive to the
electroproduction 
amplitudes 
on the nucleon for a wide range of photon virtualities. 
This may provide information on resonance transition 
form factors complementary to the information obtained from 
pion electroproduction experiments.  

\section{Single spin asymmetries in elastic electron-nucleon scattering}
\label{sec:ssa}

\begin{figure}[h]
\begin{center}
\includegraphics[width=6cm]{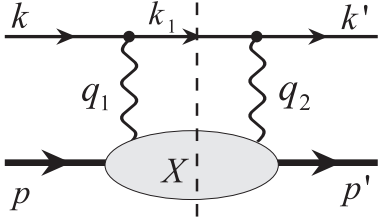}
\end{center}
\caption{The $2\gamma$ exchange diagram. 
The 
blob represents the response
of the nucleon to the scattering of the virtual photon.}
\label{fig:2gamma}
\end{figure}

The target spin asymmetries in elastic 
electron-nucleon scattering is defined by

\begin{eqnarray}
A_n \,=\, 
\frac{\sigma_\uparrow-\sigma_\downarrow}{\sigma_\uparrow+\sigma_\downarrow}\,,
\label{eq:tasymm}
\end{eqnarray} 
where $\sigma_\uparrow$ ($\sigma_\downarrow$) denotes   the 
cross section for an unpolarized beam and for a nucleon spin 
parallel (anti-parallel) to the normal polarization vector 
$\vec S_n=(\vec k\times\vec k')/|\vec k\times \vec k'|$ (with $\vec k$ and
$\vec k'$ the three-momenta of the initial and final electron, respectively).
Analogous expression as in Eq.~(\ref{eq:tasymm}) holds for the beam  
spin asymmetry ($B_n$) when we interpret $\sigma_\uparrow$ 
($\sigma_\downarrow$)
as the cross section for an unpolarized target and 
for an electron beam spin parallel (antiparallel) to the normal 
polarization vector.
As has been shown by de Rujula {\it et al.} \cite{RKR71}, 
the target and beam normal spin asymmetry 
are related to the absorptive part of the elastic $e N$ scattering
amplitude.
Since the one-photon exchange amplitude 
is purely real, the leading contribution to SSA's is of order
$O(e^2)$, and is due to an interference between one- and two-photon
exchange amplitudes, i.e.
\begin{equation}
{\rm SSA}\;=\;
\frac{2 \, {\rm Im}(\sum_{spins}T_{1\gamma}^*\cdot {\rm Abs} \,T_{2\gamma})}
{\sum_{spins}|T_{1\gamma}|^2}\, , 
\label{eq:an1}
\end{equation}
where $T_{1\gamma}$ is  the one-photon exchange amplitude, and 
$ {\rm Abs} \,T_{2\gamma}$  is the absorptive part 
of the doubly virtual Compton scattering tensor on the nucleon, 
as shown in Fig.~\ref{fig:2gamma}.
\newline
\noindent
Eq.~(\ref{eq:an1}) can be expressed in terms of a
3-dimensional phase-space integral as
\begin{eqnarray}
{\rm SSA} &=& -\frac{1}{(2\pi)^3}\frac{e^2Q^2}{D(s,Q^2)}
\int_{M^2}^{(\sqrt{s}-m_e)^2}dW^2\,\frac{|\vec{k}_1|}{4 \, \sqrt{s} }
\nonumber\\
& &\times
\int d\Omega_{k_1}\frac{1}{Q_1^2 \, Q_2^2}\,{\rm Im}
\left\{L_{\alpha\mu\nu} \, H^{\alpha\mu\nu}\right\}\,,
\label{eq:an2}
\end{eqnarray} 
where $W^2=p_X^2$ is the squared invariant mass of the intermediate state $X$,
and $s=(p+k)^2.$ 
In Eq.~(\ref{eq:an2}),  
 the momenta are defined as in Fig.~\ref{fig:2gamma},
$Q^2_1\equiv-q^2_1$ and $Q^2_2\equiv -q_2^2$ correspond with the 
virtualities of the two spacelike photons, 
$D(s, Q^2)\equiv Q^4/e^4\sum_{spins}|T_{1\gamma}|^2,$ 
and 
$L_{\alpha\mu\nu}$
and   $H^{\alpha\mu\nu}$ are the leptonic and hadronic tensors, respectively.
Furthermore, Eq.~(\ref{eq:an2}) reduces to the target or beam asymmetry 
once we specify 
the helicities for the polarized particles and take the sum over the 
helicities
of the unpolarized particles.
The 
explicit expression for 
the 
tensor  $H^{\alpha\mu\nu}$  is given by~:
\beqn
H^{\alpha\mu\nu}&=& W^{\mu\nu}\cdot
\left[\bar{u}(p',\lambda_N')\Gamma^\alpha(p', p) u(p,\lambda_N)\right]^* \, ,
\label{eq:hadt}
\eeqn
where 
$
\Gamma^\alpha(p', p) $ is the elastic photon-nucleon vertex
and $W^{\mu\nu}$  corresponds with the 
 absorptive part
of the doubly virtual Compton scattering tensor
with two {\it spacelike} photons.
The latter is given by
\begin{eqnarray}
& &W^{\mu\nu}(p',\lambda_N';p,\lambda_N)\;=\;
\sum_X \,(2\pi)^4 \, \delta^4(p+q_1-p_X)\nonumber\\
&&\times
<p'\, \lambda_N' |J^{\dagger \mu}(0)|X> \, <X|J^\nu(0)|p \, \lambda_N>\,,
\label{eq:wtensor}
\end{eqnarray}
\noindent
where the sum goes over all possible {\it on-shell} intermediate hadronic 
states $X$.
The number of intermediate states $X$ which one considers in the 
calculation sets a limit on how high in energy one can 
reliably calculate the hadronic tensor in Eq.~(\ref{eq:wtensor}). 
In addition to the elastic contribution
 ($X = N$)  which is exactly 
calculable in terms of on-shell nucleon electromagnetic 
form factors, we approximate the remaining inelastic part 
of $W^{\mu\nu}$ with
a sum over all $\pi N$ intermediate states (i.e. $X = \pi N$
in the blob of Fig.~\ref{fig:2gamma} ). 
The calculation is performed by using the unitarity relation
which allows to express $W^{\mu\nu}$
in terms of electroabsorption amplitudes $\gamma^* N \to X$ 
at different photon virtualities. 
To estimate the pion electroproduction amplitudes we use 
the phenomenological MAID analysis (version 2000)~\cite{Dre99}, 
which contains both resonant 
and non-resonant pion production mechanisms.
This same strategy has been used before in the description of 
real and virtual Compton scattering in the resonance region, 
and checked against data in Ref.~\cite{DPV03}.

\section{Results and discussion}
\label{sec:results}

In this section we show our results for both beam and target normal spin 
asymmetries for elastic electron-proton scattering. 
Our calculation covers the whole resonance region, and addresses  
measurements performed or in progress 
at MIT-Bates~\cite{Wells01}, MAMI~\cite{Maas03}, JLab~\cite{Happex,G0},
and SLAC~\cite{E158}.
%
%
\begin{figure}
\begin{center}
\includegraphics[width=7.8cm]{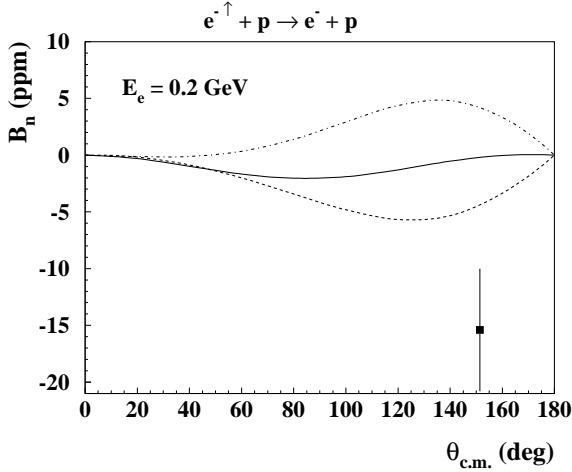}
\end{center}
\caption{Beam normal spin asymmetry $B_n$ for $e^{- \uparrow} p \to e^- p$ 
at a beam energy $E_e = 0.2$~GeV 
as function of the $c.m.$ scattering angle,  
for different hadronic intermediate states ($X$) in the blob  of
Fig.~\ref{fig:2gamma} ~: 
$N$ (dashed curve),  
$\pi N$  (dashed-dotted curve), 
sum of the $N$ and $\pi N$ (solid curve). 
The data point is from the SAMPLE Collaboration (MIT-Bates)~\cite{Wells01}.}
\label{fig:bn_mit}
\end{figure}
\begin{figure}[t]
\begin{center}
\includegraphics[width=8.3cm]{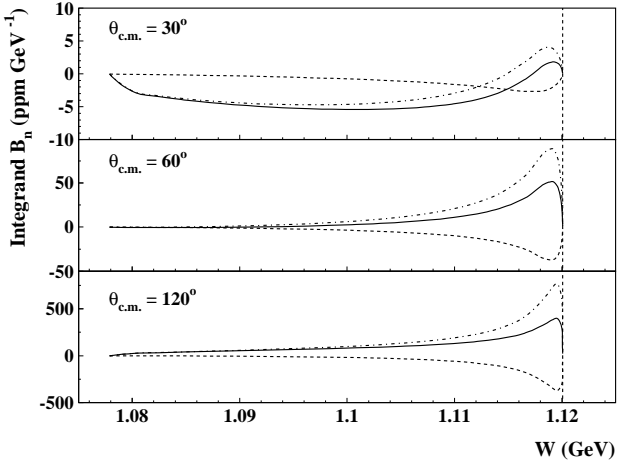}
\end{center}
\caption{
Integrand in $W$ of the beam normal spin asymmetry $B_n$ 
for $e^{- \uparrow} p \to e^- p$ 
at a beam energy of $E_e=0.2$ GeV and at different $c.m.$ scattering angles 
as indicated on the figure.
The dashed curves are the contribution from the $\pi^0 p$ channel,
the dashed-dotted curves show the contribution from the $\pi^+ n$ 
channel, and the solid curves are the sum of the contributions from the 
$\pi^+ n$ and $\pi^0 p$ channels. 
The vertical dashed line indicates the upper limit of the $W$ integration, 
i.e. $W_{max} = \sqrt{s} - m_e$.}
\label{fig:int_bn1}
\end{figure}
In Fig.~\ref{fig:bn_mit}, we show the beam normal spin asymmetry $B_n$ 
for elastic $e^{- \uparrow} p \to e^- p$ scattering at a low beam energy of 
$E_e = 0.2$~GeV. At this energy, the elastic contribution is sizeable. 
The inelastic contribution is dominated by the region of threshold pion 
production, 
as is shown in Fig.~\ref{fig:int_bn1}, where we display the integrand 
of the $W$-integration for $B_n$. When integrating the full curve in 
Fig.~\ref{fig:int_bn1} over $W$, one obtains the total inelastic contribution 
to $B_n$ (i.e. dashed-dotted curve in Fig.~\ref{fig:bn_mit}).  
The present calculation (MAID) of the threshold pion electroproduction 
is
consistent 
with chiral symmetry predictions, and is therefore largely model independent.
One notices 
that at backward {\it c.m.} angles 
(i.e. with increasing $Q^2$) the $\pi^+ n$ and $\pi^0 p$ intermediate 
states contribute with opposite sign. 
The peaked structure at the maximum 
possible value of the integration range in $W$, 
i.e. $W_{max} = \sqrt{s} - m_e$, is due to the near singularity 
(in the electron mass) 
corresponding with quasi-real Compton scattering (RCS), in which both photons in the $2\gamma$ exchange 
process become quasi-real.
This contribution at large $W$ mainly drives the results 
for the inelastic part of the beam asymmetry.
Furthermore, 
it is seen from Fig.~\ref{fig:bn_mit} that the inelastic 
and elastic contributions at a low energy of 0.2 GeV have opposite sign, 
resulting in quite a small asymmetry.
It is somewhat puzzling that the only experimental data point at this energy 
indicates a larger negative value at backward angles, 
although with quite large error bar. 
\begin{figure}[t]
\begin{center}
\includegraphics[width=8.3cm]{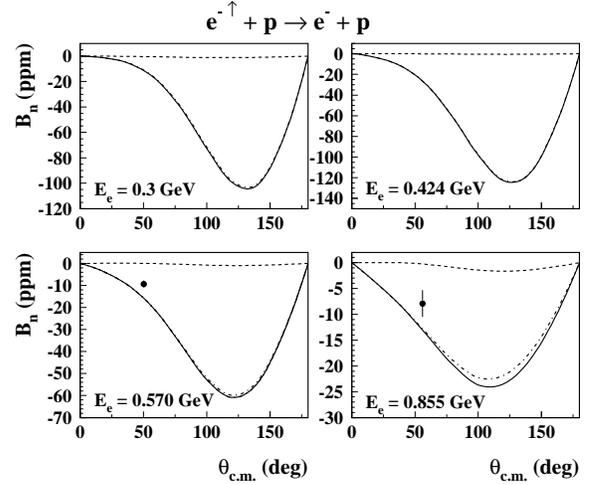}
\end{center}
\caption{Beam normal spin asymmetry $B_n$ 
for $e^{- \uparrow} p \to e^- p$ 
as function of the $c.m.$ scattering angle  
at different beam energies, as indicated on the figure.  
The meaning of the different lines is the same as in Fig.~\ref{fig:bn_mit}.
%
The data points are from the 
A4 Collaboration (MAMI)~\cite{Maas03}.}
\label{fig:bn_mami}
\end{figure}
\newline
\indent
In Fig.~\ref{fig:bn_mami}, we show $B_n$ at different beam energies below 
$E_e = 1$~GeV. It is clearly seen that at energies $E_e = 0.3$~GeV and higher  
the elastic contribution yields only a very small relative 
contribution. Therefore $B_n$ is a direct measure of the inelastic part which 
gives rise to sizeable large asymmetries, of the order of several tens of ppm 
in the backward angular range.
At forward angles, the size of the predicted  
asymmetries is compatible with the first 
high precision measurements performed at MAMI. It will be worthwhile to 
investigate if the slight overprediction (in absolute value) of $B_n$, 
in particular at $E_e = 0.57$~GeV, is also seen in a backward angle 
measurement, which is planned in the near future at MAMI. 
\begin{figure}
\begin{center}
\includegraphics[width=8.3cm]{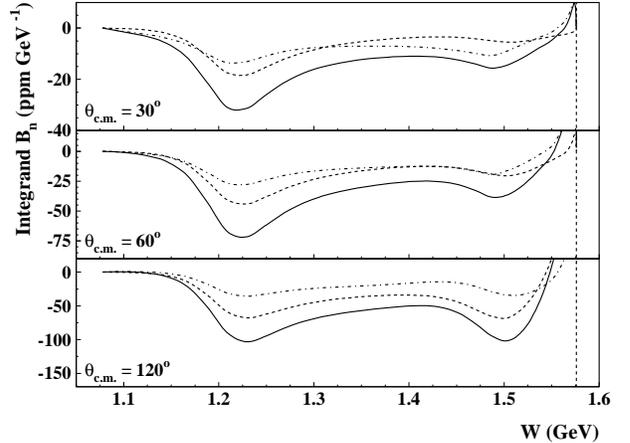}
\end{center}
\caption{
Integrand in $W$ of the beam normal spin asymmetry $B_n$ 
for $e^{- \uparrow} p \to e^- p$ 
at a beam energy of $E_e=0.855$ GeV and at different $c.m.$ 
scattering angles as indicated on the figure.
The meaning of the different lines is the same as in Fig.~\ref{fig:int_bn1}.
}
\label{fig:int_bn2}
\end{figure}
\newline
\indent
To gain a better understanding of how the inelastic 
contribution to $B_n$ arises, 
we show in Fig.~\ref{fig:int_bn2} the integrand of $B_n$ 
at $E_e = 0.855$~GeV and at different scattering angles. The resonance 
structure is clearly reflected in the integrands for both 
$\pi^+ n$ and $\pi^0 p$ channels. At forward angles, the quasi-RCS
at the endpoint $W = W_{max}$ only yields a very 
small contribution, which grows larger when going to backward angles. 
This quasi-RCS contribution is of opposite sign as the 
remainder of the integrand, and therefore determines the position of the 
maximum (absolute) value of $B_n$ when going to backward angles. 
\newline
\indent
\begin{figure}[t]
\begin{center}
\includegraphics[width=8.3cm]{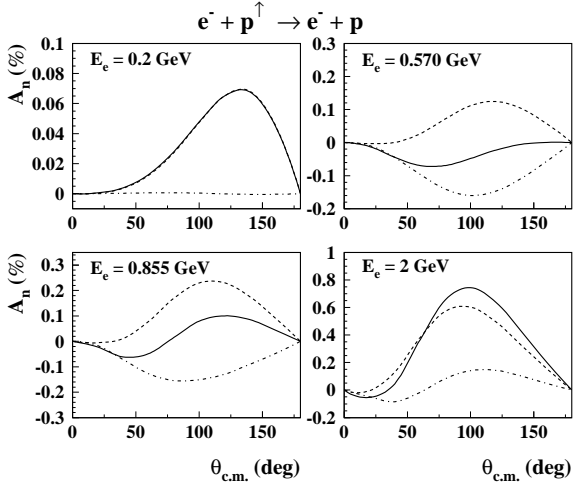}
\end{center}
\caption{Target normal spin asymmetry $A_n$ for $e^- p^\uparrow \to e^- p$
as function of the $c.m.$ scattering angle 
at different beam energies, as indicated on the figure.  
The meaning of the different lines is the same as in Fig.~\ref{fig:bn_mit}.
}
\label{fig:an_endep}
\end{figure}
We next discuss the target normal spin asymmetry $A_n$.
In Fig.~\ref{fig:an_endep}, we show the results for 
both elastic and inelastic contributions to $A_n$ at different 
beam energies. At a low beam energy of $E_e = 0.2$~GeV, $A_n$ is 
completely dominated by the elastic contribution. Going to higher beam 
energies, the inelastic contribution becomes of comparable magnitude 
as the elastic one. This is in contrast with the situation for $B_n$ 
where the elastic contribution already becomes negligible for beam 
energies around $E_e = 0.3$~GeV. 
The integrand of the inelastic contribution at a beam energy of 
$E_e = 0.855$~GeV is shown in Fig.~\ref{fig:int_an1}. The 
total inelastic result displays a $\pi^+ n$ threshold region contribution 
and a peak at the $\Delta(1232)$ resonance. Notice that the higher resonance 
region is suppressed in comparison with the corresponding integrand 
for $B_n$. Also the quasi-RCS peak around the maximum  
$W$ value is absent. 
As a result, the elastic contribution 
to $A_n$ can be of comparable magnitude as the inelastic one.
Due to the partial 
cancellation between elastic and inelastic contributions, $A_n$ 
is significantly reduced for the proton, 
taking on values around or below 0.1 \% for beam energies below 1~GeV. 
\begin{figure}
\begin{center}
\includegraphics[width=8.3cm]{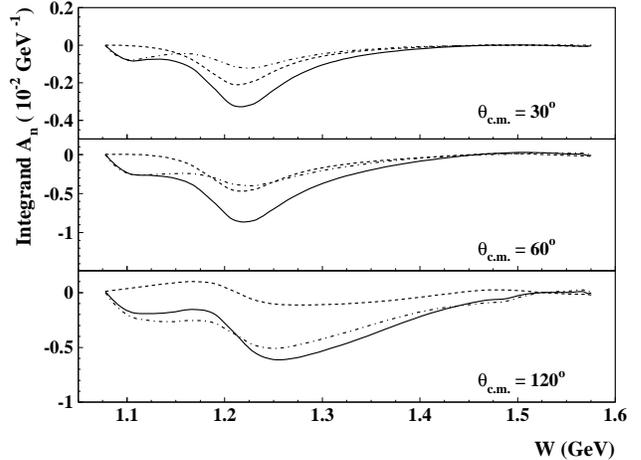}
\end{center}
\caption{
Integrand in $W$ of the target normal spin asymmetry $A_n$ 
for $e^- p^\uparrow \to e^- p$
for a beam energy of $E_e=0.855$ GeV and at different $c.m.$ scattering 
angles as indicated on the figure. 
The meaning of the different lines is the same as in Fig.~\ref{fig:int_bn1}.}
\label{fig:int_an1}
\end{figure}

\section{Conclusions}
\label{sec:concl}

In this contribution, we presented calculations for 
beam and target normal SSAs in the kinematics where several experiments 
are performed or in progress.
\newline
\noindent
Besides providing estimates for ongoing experiments, this work 
can be considered as a first step in the construction of a 
dispersion formalism for elastic electron-nucleon scattering amplitudes. 
In such a formalism, one needs a precise knowledge of the imaginary part 
as input in order to construct the real part as a dispersion integral 
over this imaginary part. The   
real part of the two-photon exchange amplitudes may yield  
corrections to elastic electron-nucleon scattering observables, 
such as the unpolarized cross sections or double polarization observables. 
Therefore it is of primary importance to quantify this piece of information, 
in order 
to increase the precision in the extraction of hadron structure quantities
such as the nucleon form factors. 
\newline
\indent


%
%

\end{document}